\documentclass[twocolumn,superscriptaddress,showpacs]{revtex4}

\usepackage{graphicx}
\usepackage{amsmath}
\usepackage{amssymb}
\usepackage{hyperref}
\usepackage{epstopdf}

\begin{document}

\title{Decohering localized waves}

\date{\today}

\author{Kristian Rayanov}
\affiliation{Max Planck Institute for the Physics of Complex Systems, N\"othnitzer Stra\ss e 38, D-01187 Dresden, Germany}
\affiliation{Institute of Physics, Chemnitz University of Technology, D-09107 Chemnitz, Germany}
\affiliation{Centre for Theoretical Chemistry and Physics, New Zealand Institute for Advanced Study, Massey University, Auckland 0745, New Zealand}
\author{G\"unter Radons}
\affiliation{Institute of Physics, Chemnitz University of Technology, D-09107 Chemnitz, Germany}
\author{Sergej Flach}
\affiliation{Max Planck Institute for the Physics of Complex Systems, N\"othnitzer Stra\ss e 38, D-01187 Dresden, Germany}
\affiliation{Centre for Theoretical Chemistry and Physics, New Zealand Institute for Advanced Study, Massey University, Auckland 0745, New Zealand}



\begin{abstract}
In the absence of confinement localization of waves takes place due to randomness or nonlinearity and relies on their phase coherence. 
We quantitatively probe the sensitivity of localized wave packets to random phase fluctuations and confirm the necessity of phase coherence for localization.
Decoherence resulting from a dynamical random environment leads to diffusive spreading and destroys linear and nonlinear localization.
We find that maximal spreading is achieved for optimal phase fluctuation characteristics 
which is a consequence of the competition between diffusion due to decoherence and ballistic transport within the mean free path distance.
\end{abstract}

\maketitle

\parskip 0pt
\section{Introduction}

Ever since the absence of diffusion of a quantum particle due to randomness on a lattice was predicted by P.W. Anderson \cite{anderson}, wave localization has become an intensively studied phenomenon.
Of particular interest is the wave character of the quantum objects which localize.
In contrast to particles waves are fields characterized by amplitude and phase, and can tunnel under a potential barrier but also be back-scattered above a given barrier.
In the absence of confinement wave localization implies that the wave does not escape to infinity even if it was energetically allowed.
Consequently Anderson localization is an interference phenomenon which relies on phase coherence.

Localization on a lattice also occurs as result of an applied DC field where it leads to Bloch oscillations \cite{wannierstarkoriginal,wannierfukuyama}.
It can even show up in translationally invariant lattices in the form of discrete breathers as a result of nonlinear interactions \cite{takeno,flachrep}.
Another source for wave localization can be a quasi-periodic potential interpolating between uncorrelated disorder and perfect periodicity.
In this case localization has been predicted by Aubry and Andr\'e \cite{aubryandre} only when  the strength of the quasi-periodic potential exceeds a certain critical value.
The latter case is largely debated.
Initially the additional importance of high enough incommensurability was stressed \cite{bellissard_physrevlett_49,ostlund_physrevlett_50,bellissard_communmath_88}.
More recently, the occurrence of a metal-insulator transition has even led to the claim that localization in the Aubry-Andr\'e model has particle character \cite{particles},
and therefore the phase coherence should not matter.

In this work we probe linear and nonlinear localization through its sensitivity to phase coherence. 
We find that loss of phase coherence directly relates to delocalization.
In experiments decoherence generally arises due to random temporary fluctuations as a result of inevitable coupling of the ideal system to its environment \cite{zurek,randomenv}.
Therefore our considerations are also related to the conceptual understanding of the effects of decoherence which is assumed to play a key role in the 
translation of the quantum world to the ``classical'' picture obtained through generic measurement processes \cite{schlosshauer}.

The importance of minimizing fluctuations becomes evident from the decade-long difficulties of experimentally observing Anderson localization.
In disordered electronic systems, where it was initially predicted, localization is hindered by phonon and electron assisted variable range hopping \cite{rangehopping,electronreview}.
Anderson localization has also been related to the study of quantum chaos in the quantum kicked rotor system \cite{AndersonQKR}.
Here, classical chaotic diffusion is suppressed by quantum interferences and leads to localization in momentum space. 
In this system the destructive impact of noise and decoherence on Anderson localization has been discussed as well, both theoretically \cite{noiseQKR} and experimentally with ultracold atoms in a pulsed standing light wave \cite{expdynamicallocalization,expnoiseQKR}. 
However, direct observation of Anderson localization of matter waves has been achieved only recently \cite{naturebilly,natureroati} with the creation of Bose-Einstein condensates where fluctuations are sufficiently reduced. 
Apart from quantum systems, the wave character of Anderson localization, with the inherent role of coherence, has also been demonstrated with light waves in coupled optical wave guides \cite{lahini}.

In the following we will investigate wave packets which are localized not only as a consequence of random disorder, but also of quasi-periodic disorder and DC fields (linear localization) and of interactions (nonlinear localization). 
We will show that specifically fluctuations in the phases, resulting from coupling to a random environment, lead to delocalization which is directly related to the loss of coherence.
The loss of localization is observed as a generic onset of a diffusive spreading regime.
Interestingly, maximal spreading and therefore delocalization is not achieved for the strongest or most frequent dephasing.
There exists an optimal rate and strength of dephasing which maximizes the wave packet's extent at a given time.
The optimal rate can be derived from knowledge of the localization volume of a given wave packet. 
All results also remain true for the case of quasi-periodic disorder, therefore localization in the Aubry-Andr\'e model has, as in the Anderson localization problem, pure wave character and in the case of matter waves is of quantum origin.

\section{Localized wave models}

We consider the model of the one-dimensional discrete nonlinear Schr\"{o}dinger equation
\begin{equation}
i \frac{\partial}{\partial t} \psi_l = \epsilon_l \psi_l + \psi_{l+1} + \psi_{l-1} + \beta |\psi_l|^2\psi_l,
\label{eq:dse}
\end{equation}
where $\psi_l$ is a complex field at site $l$ with on-site energy $\epsilon_l$ and nonlinearity strength $\beta$.
It can be derived from the Hamiltonian 
\begin{equation}
 \mathcal{H} = \sum \limits_l \epsilon_l |\psi_l|^2 + (\psi_{l+1} \psi_l^* + \psi_{l+1}^* \psi_l) + \dfrac{\beta}{2} |\psi_l|^4 
 \label{eq:hamiltonian}
\end{equation}
by $\dot{\psi_l} = \partial \mathcal{H} / \partial (i \psi_l^*)$.
Varying the total norm $S := \sum_l |\psi_l|^2$ is equivalent to varying $\beta$, therefore it is always possible to fix $S = 1$.
Then $|\psi_l|^2$ can be identified as the norm density on site $l$.
Both the energy $\mathcal{H}$ and the norm $S$ are integrals of motion. 
The discrete nonlinear Schr\"{o}dinger equation allows for the investigation of all localization phenomena described above.

Linear localization of wave packets is obtained for $\beta = 0$ and suitable potentials $\epsilon_l \neq 0$.
The resulting set of linear differential equations may be decoupled in terms of normal modes (NMs).
Separation of variables $(\psi_l(t) = A_le^{-i\lambda t})$ leads to the eigenvalue problem
\begin{equation}
 \lambda A_l = \epsilon_l A_l + A_{l+1} + A_{l-1}
 \label{eq:evp}
\end{equation}
for the only site dependent (in general complex) amplitudes $A_l$.
The normalized eigenvectors $A_{\nu, \{l\}}$ are the NMs having eigenfrequencies $\lambda_\nu$.
For localized NMs any initial excitation containing only a finite number of them will stay localized in time, leaving the NMs oscillating with their eigenfrequency respectively but independently from each other.
As will be shown later, real amplitudes $A_{\nu, l}$ (especially including negative ones) ensure localization of a single NM.

We consider the following three cases of on-site energies which yield localized NMs.
(I) Disorder (Anderson model): the on-site energies are random and chosen uniformly from the interval $[-\frac{W}{2}, \frac{W}{2}]$, where $W$ denotes the disorder strength.
The eigenfrequencies $\lambda_\nu$ lie in the interval $[-2-\frac{W}{2},2+\frac{W}{2}]$.
The NMs decay exponentially as $A_{\nu, l} \sim e^{-l/\xi}$ with a localization length $\xi(\lambda_\nu)$.
(II) DC Field (Wannier-Stark ladder): the field $E$ determines the linear growth of the on-site energies, $\epsilon_l = El$.
The eigenfrequencies are $\lambda_{\nu} = E\nu$ and the NMs are given by Bessel functions of first kind $J_l(x)$ as $A_{\nu, l} = J_{l-\nu}(2/E)$ \cite{wannierfukuyama}.
(III) Quasiperiodic potential (Aubry-Andr\'e model): the on-site energies satisfy $\epsilon_l = \zeta \cos(2\pi\alpha l)$.
The commensurability is characterized by the irrational parameter $\alpha$ while the parameter $\zeta$ describes a relative strength, similar to $W$ in the Anderson model.
A condition for localized NMs is that $\alpha$ be as far as possible from a rational number \cite{bellissard_communmath_88}, where a standard choice is to take the inverse golden mean $\alpha = \frac{\sqrt{5}-1}{2}$ \cite{ostlund_physrevb_29}.
Then NMs are exponentially localized for $\zeta > 2$ in real space and for $\zeta < 2$ in Fourier space (self-duality).
Consequently there exists a metal-insulator transition at $\zeta = 2$ in either space.

Adding a nonlinearity in general destroys the integrability of the linear system by inducing frequency shifts which result in excitations of overlapping NMs \cite{flachspreaduniversal,flachspreadchem}.
For localized NMs small enough nonlinearities have been shown to cause delocalization with subdiffusive spreading \cite{flachspreaduniversal,laptyeva_crossover_2010,bodyfelt_nonlinear_2011}.
In contrast, a large enough nonlinearity can lead to localization in form of discrete breathers when the nonlinear frequency shifts exceed the linear eigenfrequency spectrum (self-trapping) \cite{flachrep,kopidakis_absence_2008}.
Discrete breathers are exponentially localized and their time-dependence is characterized by a single frequency $\Omega_b$ as $\psi_l(t) = A_l e^{i \Omega_b t}$.

To study nonlinear localization independent of linear localization, we consider Eq. \eqref{eq:dse} with $\epsilon_l = 0, \forall l, \beta > 0$.
Then the NMs of the underlying linear system (Eq. \eqref{eq:evp} with $\epsilon_l = 0$) are extended and localization can only be attributed to discrete breathers.
Any not self-trapped part of a wave packet will spread ballistically over the entire chain.
Note that overlap of several discrete breathers in general does not lead to new localized wave packets with a quasiperiodic time-dependence \cite{flachrep}.

To quantify the spatial extent (localization volume) of an arbitrary wave packet we compute the participation number $P = 1/\sum_l|A_l|^4$ which measures the number of considerably excited sites \cite{mirlin}, and the second moment (variance) $m_2 = \sum_l (m_1-l)^2 |A_l|^2$ which measures the average norm spread around its center $m_1 = \sum_l l |A_l|^2$.
Since $m_{2}$ has the units of a square distance, a localization volume in one dimension is defined proportional to $\sqrt{m_{2}}$ \cite{locmeasure}.

\section{Localization as a coherent effect}

In this paper we want to establish a direct relationship between coherence and localization.
The main goal is to show how temporal phase fluctuations always destroy localization.
Consequently phase coherence is necessary for localization, which is a manifestation of its crucial wave character. 
Considering localization in form of localized NMs and discrete breathers, we will first identify maximally localized wave packets with maximal coherence. 
Then we will investigate the effects of phase fluctuations and decoherence.

To characterize a wave packets' degree of coherence we will utilize the basic property of the second order complex degree of coherence $\gamma$ \cite{mandelwolf} which is closely linked to the visibility of interference fringes in related interference experiments.
The essence of second order coherence lies in the investigation of correlations between two space-time points.
It is furthermore a prerequisite for coherence of higher orders that contain correlations of more space-time points.
The second order complex degree of coherence is defined as \cite{mandelwolf}
\begin{equation}
 \gamma(l_1, l_2, \tau) := \frac{\Gamma(l_1, l_2, \tau)}{\sqrt{\Gamma(l_1, l_1, 0)} \sqrt{\Gamma(l_2, l_2, 0)}},
 \label{eq:gamma}
\end{equation}
which is a normalized version of the mutual coherence function
\begin{equation}
 \Gamma(l_1, l_2, \tau) = \lim \limits_{T \rightarrow \infty} \frac{1}{T} \int \limits_0^T \psi_1^*(t) \psi_2(t+\tau) dt.
 \label{eq:Gamma}
\end{equation}
Complete coherence is achieved for $|\gamma(l_1, l_2, \tau)| = 1$, while partial coherence corresponds to $0 < |\gamma(l_1, l_2, \tau)| < 1$ and complete incoherence to $|\gamma(l_1, l_2, \tau)| = 0$.
Note that the averaging in Eq. \eqref{eq:Gamma} assumes statistically stationary evolving fields $\psi_1, \psi_2$.

Inserting a generic solution of the form $A_l e^{i\omega t}$ with site-independent $\omega$ and time-independent amplitudes$A_l$ into Eq. \eqref{eq:gamma}, it is straightforward to obtain $|\gamma| = 1$.
Thus a single NM or discrete breather always possesses complete coherence, when $\omega$ is identified with either $\lambda_{\nu}$ or $\Omega_b$, respectively.
The phases at different sites of these coherent solutions remain locked together in time.
It can also be shown easily that a superposition of more than one NM will only be partially coherent.
Since nonlinear localization is only expected in form of discrete breathers it is in general always fully coherent.
Note that here we have not made any restrictions to the complex amplitudes $A_l$ except for time-independence.
Especially we have not requested that the solutions (e.g. the NMs) be localized.
Therefore coherence is not a sufficient condition for localization.

In order to obtain a precise necessary condition between localization and coherence let us consider the evolution of the norm density $\rho_l = \psi_l^* \psi_l = |A_l|^2$.
Norm conservation of Eq. \eqref{eq:dse} ensures that the continuity equation
\begin{equation}
 0 = \frac{\partial \rho_l}{\partial t} + (j_{l+1} - j_l),
 \label{eq:con}
\end{equation}
holds.
Inserting Eq. \eqref{eq:dse} into Eq. \eqref{eq:con} one obtains the familiar result that a phase difference in a complex field $\psi_l = |\psi_l| e^{i\varphi_l}$ leads to a norm density current 
\begin{equation}
 j_l = -2 |\psi_{l+1}| |\psi_l| \sin(\varphi_{l+1} - \varphi_l).
 \label{eq:current}
\end{equation}
For time-independent $\rho_l$ the first term in Eq. (\ref{eq:con}) has to vanish, requiring $j_l$ to be actually independent of site $l$.
Now consider that localization implies $|A_l| \rightarrow 0$ for $l \rightarrow \pm \infty$.
Then the sites with $|A_l| = 0$ demand $j_l = 0 \;\forall l$ in Eq. (\ref{eq:current}).
Thus for any localized solution $j_l$ need not only be constant in space but even strictly equal to zero.
The sites where $|A_l| \neq 0$ can only satisfy $j_l = 0$ when $\varphi_{l+1} - \varphi_l = m\pi$, $m = 0, \pm 1, \pm 2, ...$
Consequently the condition of locking together the phases of neighboring sites at differences of a multiple integer of $\pi$ is necessary for localization of a single NM and discrete breather.
This is equivalent to choosing real amplitudes $A_l$ and especially allowing negative values as well.
Since the most localized wave packets consist of either a single NM or discrete breather in a linear or nonlinear model respectively the above condition is also necessary for maximal localization.

\section{Decohering and Delocalizing}

Now we will quantitatively analyze the impact of temporal phase fluctuations which in general occur when coupling to a stochastic environment is considered. 
It is clear that for a maximally localized wave packet this will destroy the fixed phase relation of neighboring sites and lead to a local norm density current.
Then complete coherence is lost and spreading takes place.
However, the wave packet does not have to delocalize completely as long as some partial coherence remains.
We will numerically show that an initial dephasing induced loss of coherence directly relates to the loss of localization and that persistent fluctuations finally delocalize the wave packet completely.

We integrate Eq. \eqref{eq:dse} using a symplectic SABA$_1$ integrator described in \cite{skokos_delocalization_2009}.
A dynamical random environment is included by considering dynamical on-site energies $\epsilon_l \rightarrow \epsilon_l + \varepsilon_l(t)$ where $\varepsilon_l(t)$ is a random process.
It can be easily shown that $\varepsilon_l(t)$ defines a dephasing term in the equations of motion of the phases $\varphi_l$.
Numerically the dephasing is implemented by altering the phases between integration steps of the unperturbed equations of motion.

We choose two different dephasing schemes.
The first one is a complete random dephasing which allows to observe the effects of dephasing as clearly as possible.
Therefore the phases $\varphi_l$ are replaced by completely new random phases chosen uniformly from $[0, 2\pi]$ at certain times of the integration.
Between these kicks on the phases the dynamics of a wave packet is solely governed by the linear or nonlinear equations of motion.
Each kick can be considered as defining a new initial configuration for the integration where all the information of the old phases is lost.
The second scheme is a quasiperiodic dephasing, similar as in \cite{randomenv}. 
Its purpose is to probe the sensitivity of localization to only small phase fluctuations which are even not completely uncorrelated.
The new phases result from the old ones as $\varphi_l(t) = \varphi_l + b \sin(\mu_l t)$ with a time-independent frequency $\mu_l$ chosen on each site randomly and uniformly from $[0, max \{\mu_l\}]$ and a constant strength of dephasing $b$.
The frequencies $\mu_{l}$ are in general incommensurate but fixed during the integration.
This leads to uncorrelated fluctuations between different sites and temporarily correlated fluctuations on each site.
Here, the phase kicks are performed after each step of integration.
To calculate the complex degree of coherence we switch off dephasing after a certain number of kicks on the phases in order to obtain statistically stationary evolving fields.
We use the time evolution at the center of norm and one neighboring site.

\begin{figure}
 \centering 
 \includegraphics[width=.99\columnwidth]{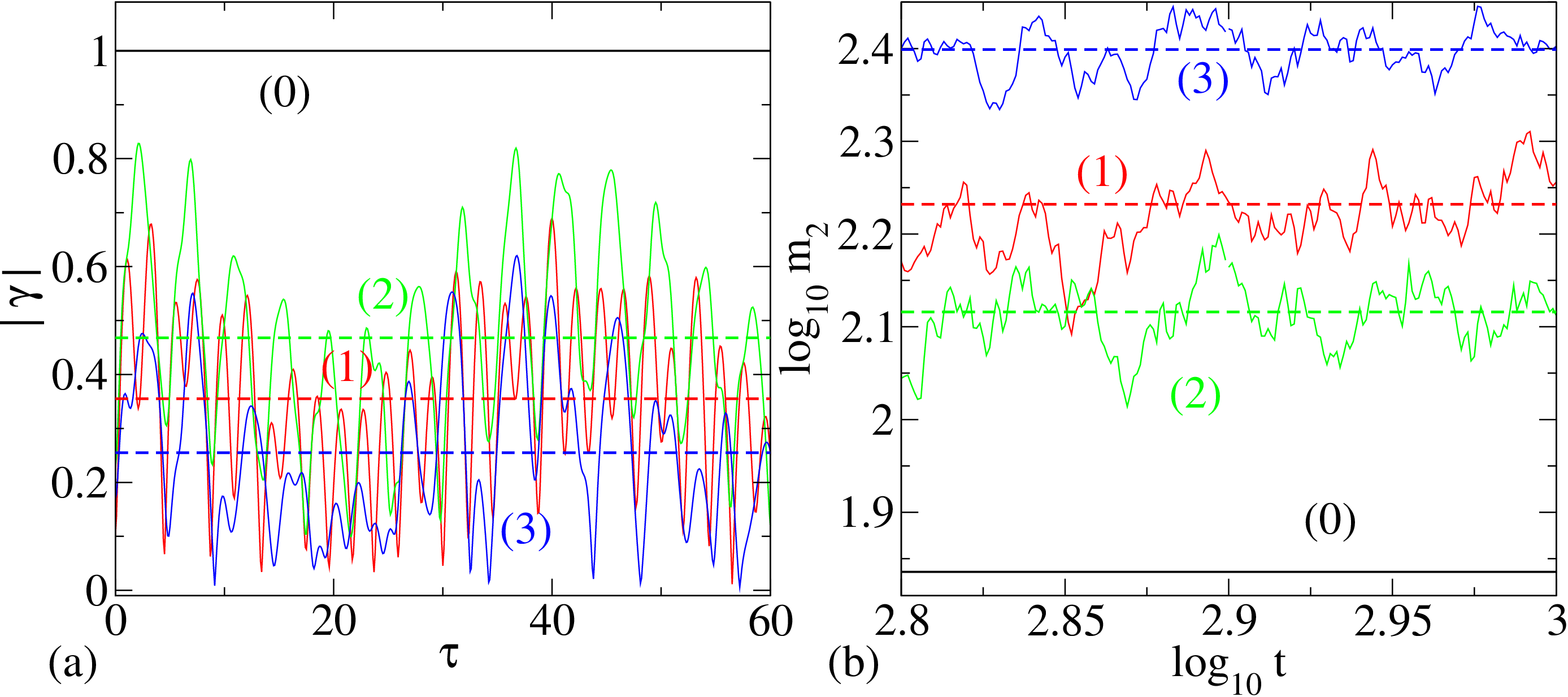}
 \caption{\scriptsize Comparison of (a) $|\gamma|$ and (b) $m_{2}$ in the Anderson model after complete random dephasing of a NM. Dephasing is switched off after a number k of kicks: k=0 (black line), 1 (red), 2 (green), 3 (blue). In (b) only times are shown when the wave packet is on average not spreading anymore. A direct correspondence between decoherence and delocalization is observed when changes in the averages of $|\gamma(\tau)|$ (dashed lines in (a)) and $m_{2}(t)$ (dashed lines in (b)) are compared.}
 \label{fig:corlin}
\end{figure}

\begin{figure}
 \centering 
 \includegraphics[width=.99\columnwidth]{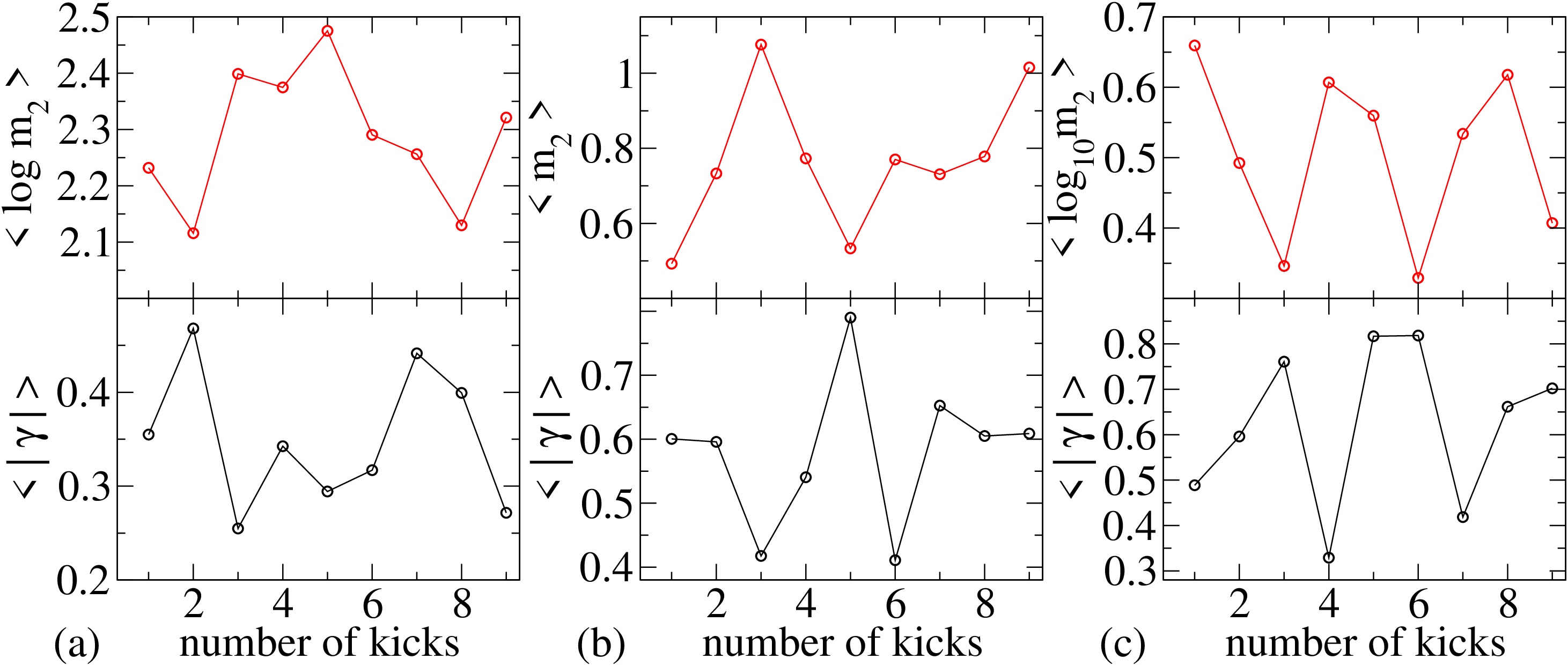}
 \caption{\scriptsize Time averages of $|\gamma|$ (lower panel), $m_{2}$ (upper panel) in the (a) Anderson model, (b) Wannier-Stark ladder, (c) Aubry-Andr\'e model when dephasing is switched off after a small number k of kicks. The averages for  k=1, 2 and 3 kicks in (a) correspond to the dashed lines in Fig.\ref{fig:corlin}. Up to a number of 8 kicks (in (a), (c), 9 kicks in (b)) an increase (decrease) in $\gamma$ corresponds to a decrease (increase) in $m_{2}$.}
 \label{fig:comparegammam2}
\end{figure}

Dephasing of a NM (Fig.\ref{fig:corlin}) shows that the decoherence is in general non-monotonic with respect to the growing number of phase kicks, however, a decrease (increase) in coherence similarly leads to a decrease (increase) in localization. 
The reason is a selective excitation and damping of overlapping NMs.
Therefore the direct correspondence between decoherence and delocalization can be found in all linear models with localized NMs (Fig.\ref{fig:comparegammam2}).
It holds for a small number of phase kicks as long as the excited NMs contain the two sites which are considered in the calculation of $|\gamma|$.

In contrast, after a few phase kicks on a discrete breather, either only a small part of the norm is radiated and the rest can again self-trap to form a new fully coherent wave packet, or the discrete breather delocalizes completely.
However, the approach to a new time-periodic trajectory of a discrete breather in phase space can take up arbitrarily long times \cite{flachrep}.
In this regime we observed seemingly localized (over accessible integration times) wave packets with only small deviations from complete coherence.
The deviations seem to be larger when the initially radiated norm (and therefore delocalization) has been larger, approximately confirming a correspondence between decoherence and delocalization.

\begin{figure}
 \centering
 \includegraphics[width=.99\columnwidth]{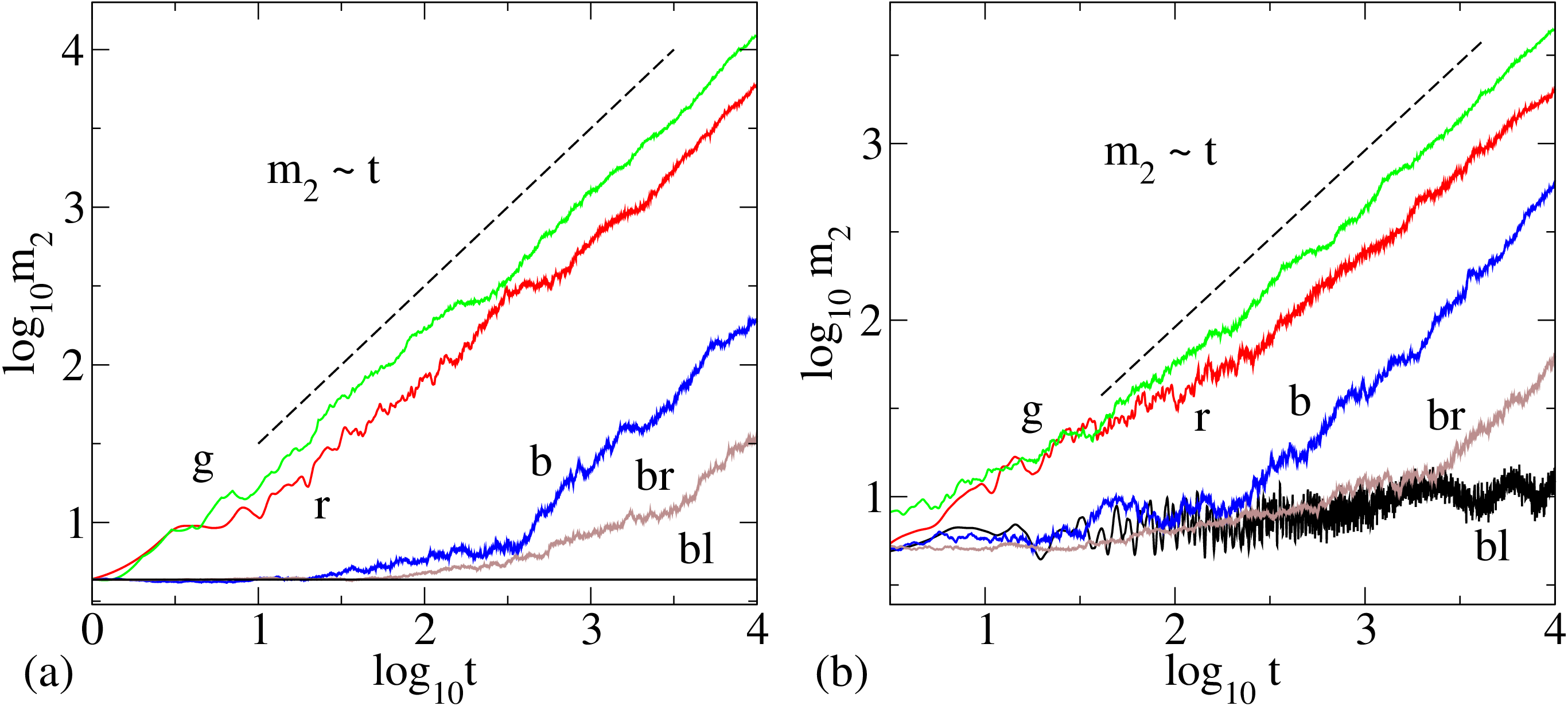}
 \caption{\scriptsize Diffusive spreading in linear chains (Anderson model) for persistent dephasing of initially one normal mode. (a) Complete random dephasing with a time $\vartriangle$ between phase kicks of $\vartriangle=10$ red (r), 1 green (g), 0.01 blue (b) and 0.001 brown (br). The black (bl) line shows the evolution without dephasing for reference. (b) Quasiperiodic dephasing with strength $b=0.0001$ black (bl), 0.0006 red (r), 0.01 green (g), 0.06 blue (b) 1.1 brown (br).}
 \label{fig:m2loglog}
\end{figure}

\begin{figure}
 \centering
 \includegraphics[width=.99\columnwidth]{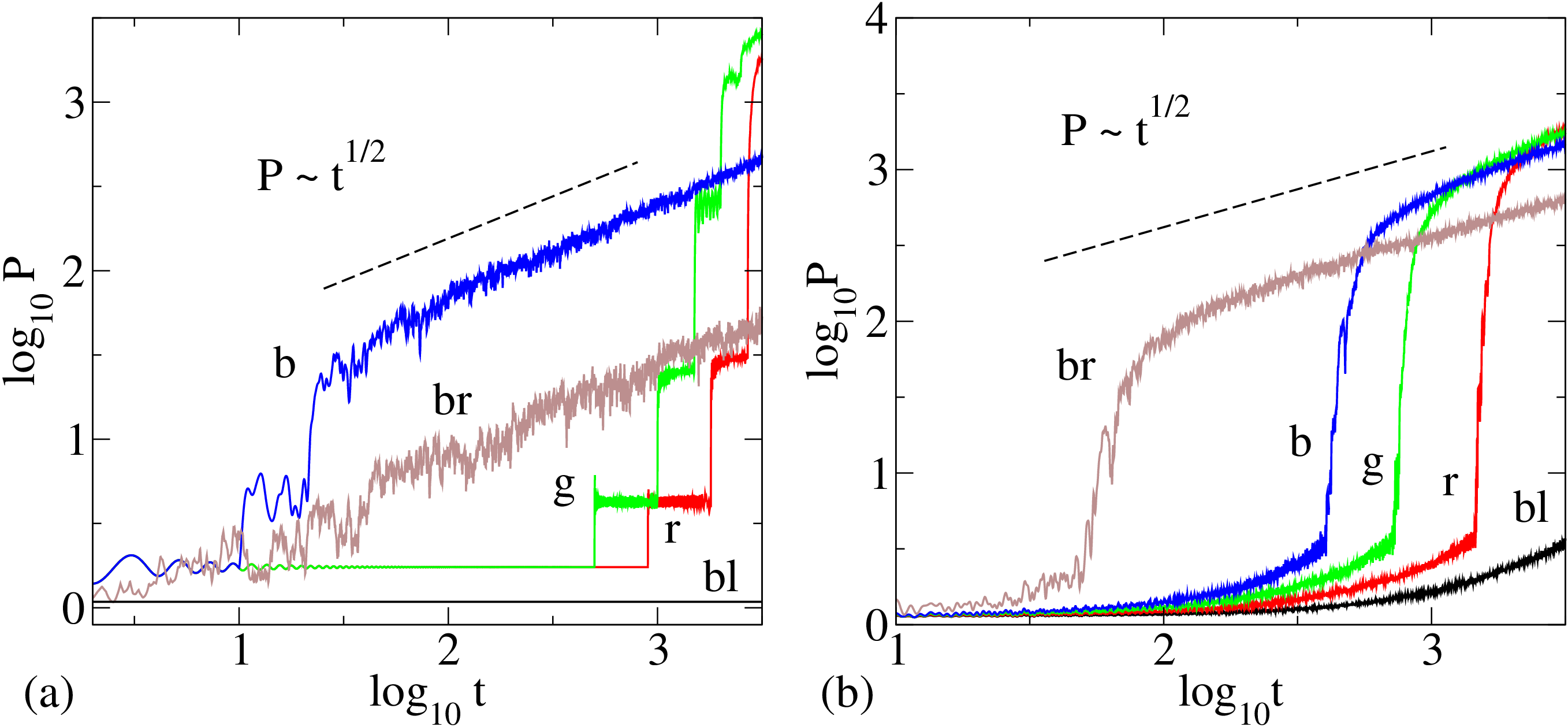}
 \caption{\scriptsize Diffusive spreading for persistent dephasing of a discrete breather. The crossover from a ballistic to diffusive spreading regime is seen as a jump-like increase of $P$. (a) Complete random dephasing with time $\vartriangle$ between phase kicks of $\vartriangle=900$ red (r), 500 green (g), 10 blue (b), 0.1 brown (br). The black (bl) line shows the evolution without dephasing for reference. (b) Quasiperiodic dephasing with strength $b=0.0001$ black (bl), 0.00015 red (r), 0.0002 green (g), 0.00025 blue (b), 0.0006 brown (br). \newline}
 \label{fig:Ploglog}
\end{figure}

A common result for both linear and nonlinear models is that coherence on average decreases for a large growing number of phase kicks.
This situation is close to an experimental one where phase fluctuations exist persistently.
We observe that decoherence by persistent dephasing leads to complete delocalization and the generic onset of a diffusive spreading regime, $m_2 \sim t$ and $P \sim \sqrt{t}$ (Fig.\ref{fig:m2loglog} and Fig.\ref{fig:Ploglog}), as can be expected for similar dynamically disordered tight-binding Hamiltonians \cite{diffusionhaken}. 
Note that the loss of norm from a discrete breather is properly characterized by the participation number.
Here, the diffusive regime is preceded by jump-like increases of $P$ which correspond to radiation of substantial parts of norm as small amplitude waves.
It is expected that discrete breathers remain robust upon radiation of small amplitude waves \cite{flachsmallamplituderadiation} which can be treated as linear background of the high amplitude breather \cite{rumpfsmallamplitude}.
Consequently, when dephasing is only performed very rarely, new self-trapping may occur leading to a series of stepwise increases of $P$ (Fig.\ref{fig:Ploglog}(a)) .

In general, the transition times before the asymptotic diffusive regime is assumed can vary.
Especially for the quasiperiodic dephasing it may be expected that even for very small $b$ normal diffusion commences after sufficiently long waiting time (see \cite{randomenv} for the Anderson model).
This shows the highly sensitive dependence of localization on phase coherence and minimal phase fluctuations.
Moreover the results do not depend on the specific range $[0, \max \mu_l]$ of dephasing frequencies $\mu_l$.
Consequently, in experiments localization is at best an intermediate regime, since arbitrarily small fluctuations can destroy localization.

Another interesting effect of random dephasing is that increasing the rate or strength of dephasing does not necessarily lead to a decrease in localization and to faster spreading.
It is rather observed that very strong and frequent dephasing even may suppress the onset of normal diffusion for an increasingly long time. 
As a  result, the wave packet extent at the final time of integration becomes maximal for a certain optimal rate and strength of dephasing. 

When the maximal extent corresponds to a diffusive spreading regime it can be attributed a maximal diffusion constant $D$.
This constant then also defines the optimal dephasing parameters for maximal delocalization at later times.
Note that in the nonlinear model (Fig.\ref{fig:Ploglog}) this is not the case.
A discrete breather is destroyed best when the onset of normal diffusion can be avoided for a long time. 
Therefore optimal rates and strengths are different for different final times considered and cannot be identified with a macroscopic constant.

\begin{figure}
 \centering
 \includegraphics[width=.99\columnwidth]{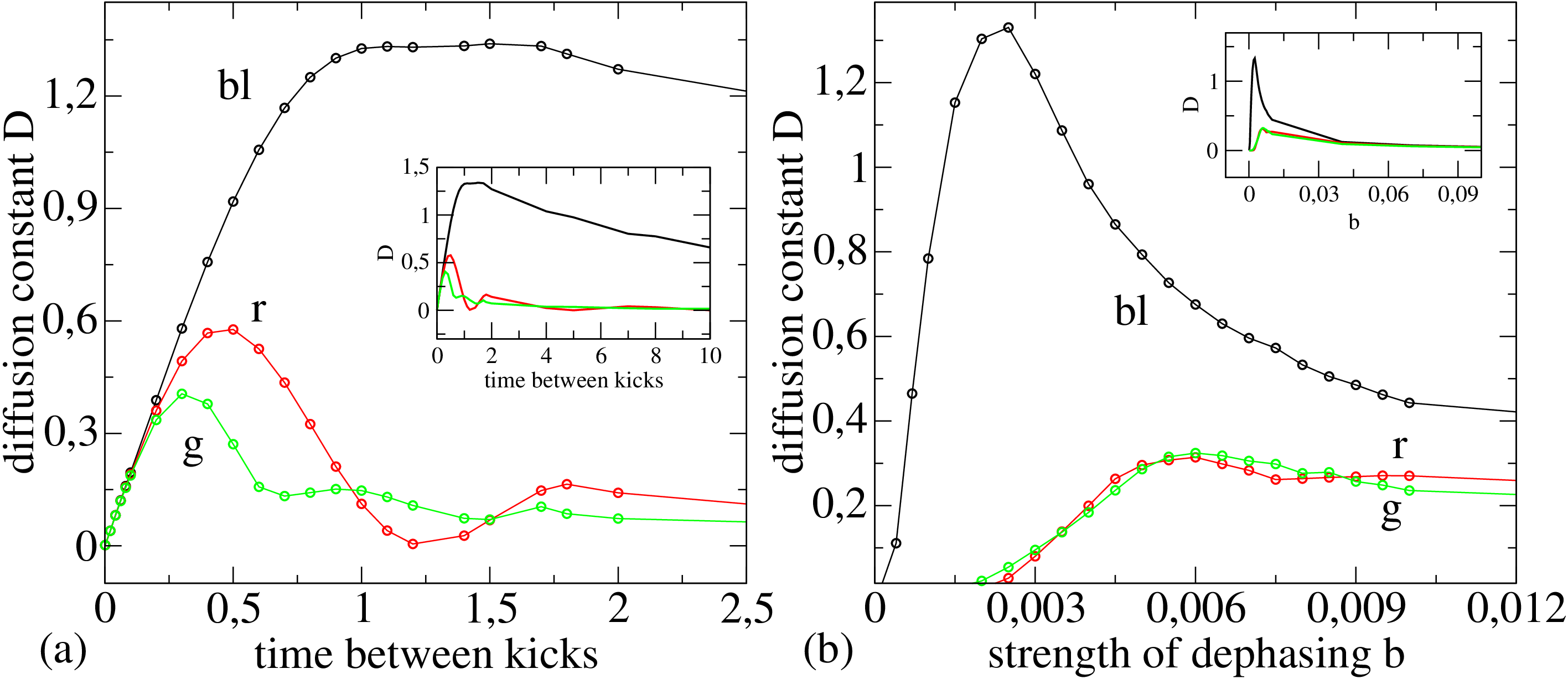}
 \caption{\scriptsize Dependence of the diffusion constant D on (a) rate and (b) strength of dephasing in the Anderson model (black lines (bl)), Wannier-Stark ladder (red lines (r)) and Aubry-Andr\'e model (green lines (g)) with the asymptotic dependence in the insets. The occurrence of a global maximum in each model is observed. Each data point of D results from linear regression of an average of $m_2(t)$ over 100 realisations of sequences of random dephasing. The connecting lines guide the eye.}
 \label{fig:Dofdept}
\end{figure}

For the linear models the diffusion constants are shown in Fig. \ref{fig:Dofdept}.  
Apart from some local maxima there clearly exists a global maximum (see also \cite{japandiffusionphysrev,japandiffusionphysica} for the Anderson model) in a wide range of rates and strengths of dephasing.
Its nature can be revealed with the following estimate in the Anderson model.
Consider that the localization length corresponds to a mean free path of quasiparticles \cite{ballvol}.
An optimal rate would leave the wave packet spread ballistically over the new accessible localization length $\xi$ before the next phase kick is applied.
Since after one kick only those NMs can be excited considerably, whose center of norm lies inside the localization volume (determined by $\xi$) of the initial wave packet, at most one localization volume becomes additionally accessible.
An upper estimate of the localization volume of a NM is of the order of $330/W^{2}$ corresponding to a localization length of $100/W^{2}$ \cite{repprogphyskramer}.
For $W = 5$ this gives a mean free path for quasiparticles of $\approx 4$.
With a maximal particle velocity (group velocity in Fourier basis) of two sites per time unit an optimal rate corresponds to one kick every two time units.
This upper estimate is in good agreement with the numerically obtained value (Fig. \ref{fig:Dofdept}(a)) of one kick per 1.6 time units. 
As a result, one could actually expect a ballistic spreading when the kicks match the times for completing the ballistic spreading into a new localization volume.
However, we always observe the onset of a diffusive regime since the new localization volumes variate while the rate of dephasing is kept constant.
In consequence of the above, optimal parameters of dephasing have to balance on the one hand decoherence and delocalization of the initial wave packet and on the other hand should not suppress too much a possibly super-diffusive spreading by imposing normal diffusion.
Therefore it becomes also clear that the optimal dephasing of a discrete breather cannot correspond to a diffusive regime since stepwise radiation of small amplitude waves allows for ballistic transport to arbitrary distances
in this case.
\vspace{-10pt}

\section{Conclusions}

In conclusion, we investigated the effects of decoherence on linear and nonlinear localized wave packets and showed that phase coherence is a necessary condition for localization.
Therefore localization, especially in the Aubry-Andr\'e model as well, is essentially a wave phenomenon and dephasing can be identified as a general mechanism of wave packet spreading. 
As soon as non-decaying random phase fluctuations occur localization is destroyed with the onset of normal diffusion. Random phase fluctuations can also occur due to deterministic chaos
and nonintegrability of nonlinear wave equations. Then  the additional dependence of the effective diffusion constant on the wave packet density results
in a slower sub-diffusion as considered in \cite{flachspreaduniversal}.
To maximize the wave packet extent at a given time there is an optimal rate and strength of dephasing resulting from a competition between quickly decohering the initial wave packet and not too much slowing down norm radiation to the exterior chain.
\vspace{-10pt}
\acknowledgements{The authors wish to thank T.V. Laptyeva and J.D. Bodyfelt for very helpful discussions.}


\begin{thebibliography}{39}
\expandafter\ifx\csname natexlab\endcsname\relax\def\natexlab#1{#1}\fi
\expandafter\ifx\csname bibnamefont\endcsname\relax
  \def\bibnamefont#1{#1}\fi
\expandafter\ifx\csname bibfnamefont\endcsname\relax
  \def\bibfnamefont#1{#1}\fi
\expandafter\ifx\csname citenamefont\endcsname\relax
  \def\citenamefont#1{#1}\fi
\expandafter\ifx\csname url\endcsname\relax
  \def\url#1{\texttt{#1}}\fi
\expandafter\ifx\csname urlprefix\endcsname\relax\def\urlprefix{URL }\fi
\providecommand{\bibinfo}[2]{#2}
\providecommand{\eprint}[2][]{\url{#2}}

\bibitem[{\citenamefont{Anderson}(1958)}]{anderson}
\bibinfo{author}{\bibfnamefont{P.~W.} \bibnamefont{Anderson}},
  \bibinfo{journal}{Phys.~Rev.} \textbf{\bibinfo{volume}{109}},
  \bibinfo{pages}{1492} (\bibinfo{year}{1958}).

\bibitem[{\citenamefont{Wannier}(1960)}]{wannierstarkoriginal}
\bibinfo{author}{\bibfnamefont{G.~H.} \bibnamefont{Wannier}},
  \bibinfo{journal}{Phys.~Rev.} \textbf{\bibinfo{volume}{117}},
  \bibinfo{pages}{432} (\bibinfo{year}{1960}).

\bibitem[{\citenamefont{Fukuyama et~al.}(1973)\citenamefont{Fukuyama, Bari, and
  Fogedby}}]{wannierfukuyama}
\bibinfo{author}{\bibfnamefont{H.}~\bibnamefont{Fukuyama}},
  \bibinfo{author}{\bibfnamefont{R.~A.} \bibnamefont{Bari}}, \bibnamefont{and}
  \bibinfo{author}{\bibfnamefont{H.~C.} \bibnamefont{Fogedby}},
  \bibinfo{journal}{Phys.~Rev.~B} \textbf{\bibinfo{volume}{8}},
  \bibinfo{pages}{5579} (\bibinfo{year}{1973}).

\bibitem[{\citenamefont{Sievers and Takeno}(1988)}]{takeno}
\bibinfo{author}{\bibfnamefont{A.~J.} \bibnamefont{Sievers}} \bibnamefont{and}
  \bibinfo{author}{\bibfnamefont{S.}~\bibnamefont{Takeno}},
  \bibinfo{journal}{Phys.~Rev.~Lett.} \textbf{\bibinfo{volume}{61}},
  \bibinfo{pages}{970} (\bibinfo{year}{1988}).

\bibitem[{\citenamefont{Flach and Gorbach}(2008)}]{flachrep}
\bibinfo{author}{\bibfnamefont{S.}~\bibnamefont{Flach}} \bibnamefont{and}
  \bibinfo{author}{\bibfnamefont{A.~V.} \bibnamefont{Gorbach}},
  \bibinfo{journal}{Phys.~Rep.} \textbf{\bibinfo{volume}{467}},
  \bibinfo{pages}{1} (\bibinfo{year}{2008}).

\bibitem[{\citenamefont{Aubry and Andr\'e}(1980)}]{aubryandre}
\bibinfo{author}{\bibfnamefont{S.}~\bibnamefont{Aubry}} \bibnamefont{and}
  \bibinfo{author}{\bibfnamefont{G.}~\bibnamefont{Andr\'e}},
  \bibinfo{journal}{Ann.~Israel~Phys.~Soc.} \textbf{\bibinfo{volume}{3}},
  \bibinfo{pages}{133} (\bibinfo{year}{1980}).

\bibitem[{\citenamefont{Bellissard et~al.}(1982)\citenamefont{Bellissard,
  Bessiss, and Moussa}}]{bellissard_physrevlett_49}
\bibinfo{author}{\bibfnamefont{J.}~\bibnamefont{Bellissard}},
  \bibinfo{author}{\bibfnamefont{D.}~\bibnamefont{Bessiss}}, \bibnamefont{and}
  \bibinfo{author}{\bibfnamefont{P.}~\bibnamefont{Moussa}},
  \bibinfo{journal}{Phys.~Rev.~Lett.} \textbf{\bibinfo{volume}{49}},
  \bibinfo{pages}{701} (\bibinfo{year}{1982}).

\bibitem[{\citenamefont{Ostlund et~al.}(1983)\citenamefont{Ostlund, Pandit,
  Rand, Schellnhuber, and Siggia}}]{ostlund_physrevlett_50}
\bibinfo{author}{\bibfnamefont{S.}~\bibnamefont{Ostlund}},
  \bibinfo{author}{\bibfnamefont{R.}~\bibnamefont{Pandit}},
  \bibinfo{author}{\bibfnamefont{D.}~\bibnamefont{Rand}},
  \bibinfo{author}{\bibfnamefont{H.~J.} \bibnamefont{Schellnhuber}},
  \bibnamefont{and} \bibinfo{author}{\bibfnamefont{E.~D.}
  \bibnamefont{Siggia}}, \bibinfo{journal}{Phys.~Rev.~Lett.}
  \textbf{\bibinfo{volume}{50}}, \bibinfo{pages}{1873} (\bibinfo{year}{1983}).

\bibitem[{\citenamefont{Bellissard et~al.}(1983)\citenamefont{Bellissard, Lima,
  and Testard}}]{bellissard_communmath_88}
\bibinfo{author}{\bibfnamefont{J.}~\bibnamefont{Bellissard}},
  \bibinfo{author}{\bibfnamefont{R.}~\bibnamefont{Lima}}, \bibnamefont{and}
  \bibinfo{author}{\bibfnamefont{D.}~\bibnamefont{Testard}},
  \bibinfo{journal}{Commun.~Math.~Phys.} \textbf{\bibinfo{volume}{88}},
  \bibinfo{pages}{207} (\bibinfo{year}{1983}).

\bibitem[{\citenamefont{Albert and Leboeuf}(2010)}]{particles}
\bibinfo{author}{\bibfnamefont{M.}~\bibnamefont{Albert}} \bibnamefont{and}
  \bibinfo{author}{\bibfnamefont{P.}~\bibnamefont{Leboeuf}},
  \bibinfo{journal}{Phys.~Rev.~A} \textbf{\bibinfo{volume}{81}},
  \bibinfo{pages}{013614} (\bibinfo{year}{2010}).

\bibitem[{\citenamefont{Zurek}(2003)}]{zurek}
\bibinfo{author}{\bibfnamefont{W.~H.} \bibnamefont{Zurek}},
  \bibinfo{journal}{Rev.~Mod.~Phys.} \textbf{\bibinfo{volume}{75}},
  \bibinfo{pages}{717} (\bibinfo{year}{2003}).

\bibitem[{\citenamefont{Yin et~al.}(2008)\citenamefont{Yin, Katsanos, and
  Evangelou}}]{randomenv}
\bibinfo{author}{\bibfnamefont{Y.}~\bibnamefont{Yin}},
  \bibinfo{author}{\bibfnamefont{D.~E.} \bibnamefont{Katsanos}},
  \bibnamefont{and} \bibinfo{author}{\bibfnamefont{S.~N.}
  \bibnamefont{Evangelou}}, \bibinfo{journal}{Phys.~Rev.~A}
  \textbf{\bibinfo{volume}{77}}, \bibinfo{pages}{022302}
  (\bibinfo{year}{2008}).

\bibitem[{\citenamefont{Schlosshauer}(2004)}]{schlosshauer}
\bibinfo{author}{\bibfnamefont{M.}~\bibnamefont{Schlosshauer}},
  \bibinfo{journal}{Rev.~Mod.~Phys.} \textbf{\bibinfo{volume}{76}},
  \bibinfo{pages}{1267} (\bibinfo{year}{2004}).

\bibitem[{\citenamefont{Marnieros et~al.}(2000)\citenamefont{Marnieros,
  Berg\'e, Juillard, and Dumoulin}}]{rangehopping}
\bibinfo{author}{\bibfnamefont{S.}~\bibnamefont{Marnieros}},
  \bibinfo{author}{\bibfnamefont{L.}~\bibnamefont{Berg\'e}},
  \bibinfo{author}{\bibfnamefont{A.}~\bibnamefont{Juillard}}, \bibnamefont{and}
  \bibinfo{author}{\bibfnamefont{L.}~\bibnamefont{Dumoulin}},
  \bibinfo{journal}{Phys.~Rev.~Lett.} \textbf{\bibinfo{volume}{84}},
  \bibinfo{pages}{2469} (\bibinfo{year}{2000}).

\bibitem[{\citenamefont{Basko et~al.}(2006)\citenamefont{Basko, Aleiner, and
  Altshuler}}]{electronreview}
\bibinfo{author}{\bibfnamefont{D.~M.} \bibnamefont{Basko}},
  \bibinfo{author}{\bibfnamefont{I.~L.} \bibnamefont{Aleiner}},
  \bibnamefont{and} \bibinfo{author}{\bibfnamefont{B.~L.}
  \bibnamefont{Altshuler}}, \bibinfo{journal}{Ann.~Phys.}
  \textbf{\bibinfo{volume}{321}}, \bibinfo{pages}{1126} (\bibinfo{year}{2006}).

\bibitem[{\citenamefont{Fishman et~al.}(1982)\citenamefont{Fishman, Grempel,
  and Prange}}]{AndersonQKR}
\bibinfo{author}{\bibfnamefont{S.}~\bibnamefont{Fishman}},
  \bibinfo{author}{\bibfnamefont{D.~R.} \bibnamefont{Grempel}},
  \bibnamefont{and} \bibinfo{author}{\bibfnamefont{R.~E.}
  \bibnamefont{Prange}}, \bibinfo{journal}{Phys.~Rev.~Lett.}
  \textbf{\bibinfo{volume}{49}}, \bibinfo{pages}{509} (\bibinfo{year}{1982}).

\bibitem[{\citenamefont{Ott et~al.}(1984)\citenamefont{Ott, Antonsen, and
  Hanson}}]{noiseQKR}
\bibinfo{author}{\bibfnamefont{E.}~\bibnamefont{Ott}},
  \bibinfo{author}{\bibfnamefont{T.~M.} \bibnamefont{Antonsen}},
  \bibnamefont{and} \bibinfo{author}{\bibfnamefont{J.~D.}
  \bibnamefont{Hanson}}, \bibinfo{journal}{Phys.~Rev.~Lett.}
  \textbf{\bibinfo{volume}{53}}, \bibinfo{pages}{2187} (\bibinfo{year}{1984}).

\bibitem[{\citenamefont{Moore et~al.}(1994)\citenamefont{Moore, C., Bharucha,
  Williams, and Raizen}}]{expdynamicallocalization}
\bibinfo{author}{\bibfnamefont{F.~L.} \bibnamefont{Moore}},
  \bibinfo{author}{\bibfnamefont{R.~J.} \bibnamefont{C.}},
  \bibinfo{author}{\bibfnamefont{C.}~\bibnamefont{Bharucha}},
  \bibinfo{author}{\bibfnamefont{P.~E.} \bibnamefont{Williams}},
  \bibnamefont{and} \bibinfo{author}{\bibfnamefont{M.~G.}
  \bibnamefont{Raizen}}, \bibinfo{journal}{Phys.~Rev.~Lett.}
  \textbf{\bibinfo{volume}{73}}, \bibinfo{pages}{2974} (\bibinfo{year}{1994}).

\bibitem[{\citenamefont{Ammann et~al.}(1998)\citenamefont{Ammann, Gray,
  Shvarchuck, and Christensen}}]{expnoiseQKR}
\bibinfo{author}{\bibfnamefont{H.}~\bibnamefont{Ammann}},
  \bibinfo{author}{\bibfnamefont{R.}~\bibnamefont{Gray}},
  \bibinfo{author}{\bibfnamefont{I.}~\bibnamefont{Shvarchuck}},
  \bibnamefont{and}
  \bibinfo{author}{\bibfnamefont{N.}~\bibnamefont{Christensen}},
  \bibinfo{journal}{Phys.~Rev.~Lett.} \textbf{\bibinfo{volume}{80}},
  \bibinfo{pages}{4111} (\bibinfo{year}{1998}).

\bibitem[{\citenamefont{Billy et~al.}(2008)\citenamefont{Billy, Josse, Zuo,
  Bernard, Hambrecht, Lugan, Cl\'ement, Sanchez-Palencia, Bouyer, and
  Aspect}}]{naturebilly}
\bibinfo{author}{\bibfnamefont{J.}~\bibnamefont{Billy}},
  \bibinfo{author}{\bibfnamefont{V.}~\bibnamefont{Josse}},
  \bibinfo{author}{\bibfnamefont{Z.}~\bibnamefont{Zuo}},
  \bibinfo{author}{\bibfnamefont{A.}~\bibnamefont{Bernard}},
  \bibinfo{author}{\bibfnamefont{B.}~\bibnamefont{Hambrecht}},
  \bibinfo{author}{\bibfnamefont{P.}~\bibnamefont{Lugan}},
  \bibinfo{author}{\bibfnamefont{D.}~\bibnamefont{Cl\'ement}},
  \bibinfo{author}{\bibfnamefont{L.}~\bibnamefont{Sanchez-Palencia}},
  \bibinfo{author}{\bibfnamefont{P.}~\bibnamefont{Bouyer}}, \bibnamefont{and}
  \bibinfo{author}{\bibfnamefont{A.}~\bibnamefont{Aspect}},
  \bibinfo{journal}{Nature} \textbf{\bibinfo{volume}{453}},
  \bibinfo{pages}{891} (\bibinfo{year}{2008}).

\bibitem[{\citenamefont{Roati et~al.}(2008)\citenamefont{Roati, D'Errico,
  Fallani, Fattori, Fort, Zaccanti, Modugno, Modugno, and
  Inguscio}}]{natureroati}
\bibinfo{author}{\bibfnamefont{G.}~\bibnamefont{Roati}},
  \bibinfo{author}{\bibfnamefont{C.}~\bibnamefont{D'Errico}},
  \bibinfo{author}{\bibfnamefont{L.}~\bibnamefont{Fallani}},
  \bibinfo{author}{\bibfnamefont{M.}~\bibnamefont{Fattori}},
  \bibinfo{author}{\bibfnamefont{C.}~\bibnamefont{Fort}},
  \bibinfo{author}{\bibfnamefont{M.}~\bibnamefont{Zaccanti}},
  \bibinfo{author}{\bibfnamefont{G.}~\bibnamefont{Modugno}},
  \bibinfo{author}{\bibfnamefont{M.}~\bibnamefont{Modugno}}, \bibnamefont{and}
  \bibinfo{author}{\bibfnamefont{M.}~\bibnamefont{Inguscio}},
  \bibinfo{journal}{Nature} \textbf{\bibinfo{volume}{453}},
  \bibinfo{pages}{895} (\bibinfo{year}{2008}).

\bibitem[{\citenamefont{Lahini et~al.}(2008)\citenamefont{Lahini, Avidan,
  Pozzi, Sorel, Morandotti, Christodoulides, and Silberberg}}]{lahini}
\bibinfo{author}{\bibfnamefont{Y.}~\bibnamefont{Lahini}},
  \bibinfo{author}{\bibfnamefont{A.}~\bibnamefont{Avidan}},
  \bibinfo{author}{\bibfnamefont{F.}~\bibnamefont{Pozzi}},
  \bibinfo{author}{\bibfnamefont{M.}~\bibnamefont{Sorel}},
  \bibinfo{author}{\bibfnamefont{R.}~\bibnamefont{Morandotti}},
  \bibinfo{author}{\bibfnamefont{D.~N.} \bibnamefont{Christodoulides}},
  \bibnamefont{and}
  \bibinfo{author}{\bibfnamefont{Y.}~\bibnamefont{Silberberg}},
  \bibinfo{journal}{Phys.~Rev.~Lett.} \textbf{\bibinfo{volume}{100}},
  \bibinfo{pages}{013906} (\bibinfo{year}{2008}).

\bibitem[{\citenamefont{Ostlund and Pandit}(1984)}]{ostlund_physrevb_29}
\bibinfo{author}{\bibfnamefont{S.}~\bibnamefont{Ostlund}} \bibnamefont{and}
  \bibinfo{author}{\bibfnamefont{R.}~\bibnamefont{Pandit}},
  \bibinfo{journal}{Phys.~Rev.~B} \textbf{\bibinfo{volume}{29}},
  \bibinfo{pages}{1394} (\bibinfo{year}{1984}).

\bibitem[{\citenamefont{Flach et~al.}(2009)\citenamefont{Flach, Krimer, and
  Skokos}}]{flachspreaduniversal}
\bibinfo{author}{\bibfnamefont{S.}~\bibnamefont{Flach}},
  \bibinfo{author}{\bibfnamefont{D.}~\bibnamefont{Krimer}}, \bibnamefont{and}
  \bibinfo{author}{\bibfnamefont{C.}~\bibnamefont{Skokos}},
  \bibinfo{journal}{Phys.~Rev.~Lett.} \textbf{\bibinfo{volume}{102}},
  \bibinfo{pages}{024101} (\bibinfo{year}{2009}).

\bibitem[{\citenamefont{Flach}(2010)}]{flachspreadchem}
\bibinfo{author}{\bibfnamefont{S.}~\bibnamefont{Flach}},
  \bibinfo{journal}{Chem.~Phys.} \textbf{\bibinfo{volume}{375}},
  \bibinfo{pages}{548} (\bibinfo{year}{2010}).

\bibitem[{\citenamefont{Laptyeva et~al.}(2010)\citenamefont{Laptyeva, Bodyfelt,
  Krimer, Skokos, and Flach}}]{laptyeva_crossover_2010}
\bibinfo{author}{\bibfnamefont{T.~V.} \bibnamefont{Laptyeva}},
  \bibinfo{author}{\bibfnamefont{J.~D.} \bibnamefont{Bodyfelt}},
  \bibinfo{author}{\bibfnamefont{D.~O.} \bibnamefont{Krimer}},
  \bibinfo{author}{\bibfnamefont{C.}~\bibnamefont{Skokos}}, \bibnamefont{and}
  \bibinfo{author}{\bibfnamefont{S.}~\bibnamefont{Flach}},
  \bibinfo{journal}{Europhys.~Lett.} \textbf{\bibinfo{volume}{91}},
  \bibinfo{pages}{30001} (\bibinfo{year}{2010}).

\bibitem[{\citenamefont{Bodyfelt et~al.}(2011)\citenamefont{Bodyfelt, Laptyeva,
  Skokos, Krimer, and Flach}}]{bodyfelt_nonlinear_2011}
\bibinfo{author}{\bibfnamefont{J.~D.} \bibnamefont{Bodyfelt}},
  \bibinfo{author}{\bibfnamefont{T.~V.} \bibnamefont{Laptyeva}},
  \bibinfo{author}{\bibfnamefont{C.}~\bibnamefont{Skokos}},
  \bibinfo{author}{\bibfnamefont{D.~O.} \bibnamefont{Krimer}},
  \bibnamefont{and} \bibinfo{author}{\bibfnamefont{S.}~\bibnamefont{Flach}},
  \bibinfo{journal}{Phys.~Rev.~E} \textbf{\bibinfo{volume}{84}},
  \bibinfo{pages}{016205} (\bibinfo{year}{2011}).

\bibitem[{\citenamefont{Kopidakis et~al.}(2008)\citenamefont{Kopidakis,
  Komineas, Flach, and Aubry}}]{kopidakis_absence_2008}
\bibinfo{author}{\bibfnamefont{G.}~\bibnamefont{Kopidakis}},
  \bibinfo{author}{\bibfnamefont{S.}~\bibnamefont{Komineas}},
  \bibinfo{author}{\bibfnamefont{S.}~\bibnamefont{Flach}}, \bibnamefont{and}
  \bibinfo{author}{\bibfnamefont{S.}~\bibnamefont{Aubry}},
  \bibinfo{journal}{Phys.~Rev.~Lett.} \textbf{\bibinfo{volume}{100}},
  \bibinfo{pages}{084103} (\bibinfo{year}{2008}).

\bibitem[{\citenamefont{Mirlin}(2000)}]{mirlin}
\bibinfo{author}{\bibfnamefont{A.~D.} \bibnamefont{Mirlin}},
  \bibinfo{journal}{Phys.~Rep.} \textbf{\bibinfo{volume}{326}},
  \bibinfo{pages}{259} (\bibinfo{year}{2000}).

\bibitem[{\citenamefont{Krimer and Flach}(2010)}]{locmeasure}
\bibinfo{author}{\bibfnamefont{D.}~\bibnamefont{Krimer}} \bibnamefont{and}
  \bibinfo{author}{\bibfnamefont{S.}~\bibnamefont{Flach}},
  \bibinfo{journal}{Phys.~Rev.~E} \textbf{\bibinfo{volume}{82}},
  \bibinfo{pages}{046221} (\bibinfo{year}{2010}).

\bibitem[{\citenamefont{Mandel and Wolf}(1995)}]{mandelwolf}
\bibinfo{author}{\bibfnamefont{L.}~\bibnamefont{Mandel}} \bibnamefont{and}
  \bibinfo{author}{\bibfnamefont{E.}~\bibnamefont{Wolf}},
  \emph{\bibinfo{title}{Optical Coherence and Quantum Optics}}
  (\bibinfo{publisher}{{Cambridge University Press}}, \bibinfo{address}{New
  York}, \bibinfo{year}{1995}).

\bibitem[{\citenamefont{Skokos et~al.}(2009)\citenamefont{Skokos, Krimer,
  Komineas, and Flach}}]{skokos_delocalization_2009}
\bibinfo{author}{\bibfnamefont{C.}~\bibnamefont{Skokos}},
  \bibinfo{author}{\bibfnamefont{D.}~\bibnamefont{Krimer}},
  \bibinfo{author}{\bibfnamefont{S.}~\bibnamefont{Komineas}}, \bibnamefont{and}
  \bibinfo{author}{\bibfnamefont{S.}~\bibnamefont{Flach}},
  \bibinfo{journal}{Phys.~Rev.~E} \textbf{\bibinfo{volume}{79}},
  \bibinfo{pages}{056211} (\bibinfo{year}{2009}).

\bibitem[{\citenamefont{Schwarzer and Haken}(1972)}]{diffusionhaken}
\bibinfo{author}{\bibfnamefont{E.}~\bibnamefont{Schwarzer}} \bibnamefont{and}
  \bibinfo{author}{\bibfnamefont{H.}~\bibnamefont{Haken}},
  \bibinfo{journal}{Phys.~Lett.} \textbf{\bibinfo{volume}{42A}},
  \bibinfo{pages}{317} (\bibinfo{year}{1972}).

\bibitem[{\citenamefont{Flach et~al.}(2005)\citenamefont{Flach, Fleurov, and
  Gorbach}}]{flachsmallamplituderadiation}
\bibinfo{author}{\bibfnamefont{S.}~\bibnamefont{Flach}},
  \bibinfo{author}{\bibfnamefont{V.}~\bibnamefont{Fleurov}}, \bibnamefont{and}
  \bibinfo{author}{\bibfnamefont{A.~V.} \bibnamefont{Gorbach}},
  \bibinfo{journal}{Phys.~Rev.~B} \textbf{\bibinfo{volume}{71}},
  \bibinfo{pages}{064302} (\bibinfo{year}{2005}).

\bibitem[{\citenamefont{Rumpf}(2009)}]{rumpfsmallamplitude}
\bibinfo{author}{\bibfnamefont{B.}~\bibnamefont{Rumpf}},
  \bibinfo{journal}{Physica D} \textbf{\bibinfo{volume}{238}},
  \bibinfo{pages}{2067} (\bibinfo{year}{2009}).

\bibitem[{\citenamefont{Yamada and Ikeda}(1999)}]{japandiffusionphysrev}
\bibinfo{author}{\bibfnamefont{H.}~\bibnamefont{Yamada}} \bibnamefont{and}
  \bibinfo{author}{\bibfnamefont{K.~S.} \bibnamefont{Ikeda}},
  \bibinfo{journal}{Phys.~Rev.~E} \textbf{\bibinfo{volume}{59}},
  \bibinfo{pages}{5214} (\bibinfo{year}{1999}).

\bibitem[{\citenamefont{Ezaki and Shibata}(1992)}]{japandiffusionphysica}
\bibinfo{author}{\bibfnamefont{H.}~\bibnamefont{Ezaki}} \bibnamefont{and}
  \bibinfo{author}{\bibfnamefont{F.}~\bibnamefont{Shibata}},
  \bibinfo{journal}{Physica A} \textbf{\bibinfo{volume}{187}},
  \bibinfo{pages}{267} (\bibinfo{year}{1992}).

\bibitem[{\citenamefont{Thouless}(1973)}]{ballvol}
\bibinfo{author}{\bibfnamefont{D.~J.} \bibnamefont{Thouless}},
  \bibinfo{journal}{J.~Phys.~C} \textbf{\bibinfo{volume}{6}},
  \bibinfo{pages}{L49} (\bibinfo{year}{1973}).

\bibitem[{\citenamefont{Kramer and MacKinnon}(1993)}]{repprogphyskramer}
\bibinfo{author}{\bibfnamefont{B.}~\bibnamefont{Kramer}} \bibnamefont{and}
  \bibinfo{author}{\bibfnamefont{A.}~\bibnamefont{MacKinnon}},
  \bibinfo{journal}{Rep.~Prog.~Phys.} \textbf{\bibinfo{volume}{56}},
  \bibinfo{pages}{1469} (\bibinfo{year}{1993}).

\end{thebibliography}

\end{document}